\documentclass[aps, pra, reprint, superscriptaddress,longbibliography]{revtex4-2}
\usepackage[colorlinks,bookmarks=false,citecolor=blue,linkcolor=red,urlcolor=blue]{hyperref}
\usepackage{verbatim,float}
\usepackage{color}
\usepackage{amsmath,amssymb,bm,braket,float,mathtools,subfigure}
\usepackage{graphicx,xfrac,appendix}
\usepackage[english]{babel}
\usepackage{amsmath}
\usepackage[T1]{fontenc}
\usepackage[latin9]{inputenc}
\usepackage{graphicx}
\makeatletter

\usepackage{epstopdf,float}
\usepackage{epstopdf}
\makeatother

\begin{document}
\title{High-Winding-Number Zero-Energy Edge States in Rhombohedral-Stacked Su-Schrieffer-Heeger Multilayers}

\author{Feng Lu}
\affiliation{Department of Physics, Zhejiang Normal University, Jinhua 321004, China}

\author{Ao Zhou}
\affiliation{Department of Physics, Zhejiang Normal University, Jinhua 321004, China}

\author{Shujie Cheng}
\email{chengsj@zjnu.edu.cn}
\affiliation{Xingzhi College, Zhejiang Normal University, Lanxi 321100, China}
\affiliation{Department of Physics, Zhejiang Normal University, Jinhua 321004, China}

\author{Gao Xianlong}
\email{gaoxl@zjnu.edu.cn}
\affiliation{Department of Physics, Zhejiang Normal University, Jinhua 321004, China}

\begin{abstract}
We study the topological properties of rhombohedral-stacked N-layer Su-Schrieffer-Heeger networks with interlayer coupling. We find that these systems exhibit $2N$-fold degenerate zero-energy edge states with winding number $W=N$, providing a direct route to high-winding-number topological phases where $W$ equals the layer number. Using effective Hamiltonian theory and Zak phase calculations, we demonstrate that the winding number scales linearly with $N$ through a layer-by-layer topological amplification mechanism. We introduce the Wigner entropy as a novel detection method for these edge states, showing that topological boundary states exhibit significantly enhanced Wigner entropy compared to bulk states. Our results establish rhombohedral stacking as a systematic approach for engineering high-winding-number topological insulators with potential applications in quantum information processing.
\end{abstract}

\maketitle

\section{Introduction}
Topological insulators, characterized by an insulating bulk and conducting boundary states, have revolutionized condensed-matter physics by uncovering novel quantum phenomena rooted in nontrivial band topology\cite{RevModPhys.82.3045,shen2017topological,liu2016quantum,chang2023colloquium,thouless1982quantized}. Among the diverse topological systems, the Su-Schrieffer-Heeger (SSH) model, proposed initially to describe charge density waves in polyacetylene serves as a paradigmatic one-dimensional (1D) topological platform\cite{su1979solitons}.
Its core feature lies in the interplay between nearest-neighbor ($\nu$) and next-nearest-neighbor ($\omega$) hopping amplitudes: when $|\nu| \neq |\omega|$, the system exhibits a nontrivial topological phase characterized by zero-energy edge states, whose number is dictated by the winding number of the Bloch Hamiltonian, a $\mathbb{Z}_2$-invariant for the standard SSH model\cite{su1979solitons,kivelson1987solitons}.

The SSH model's conceptual simplicity has motivated its extension into multi-dimensional and multilayered configurations \cite{PhysRevB.106.054511,PhysRevB.108.245140,PhysRevB.107.054105}. This is driven by the fact that topological phenomena in low-dimensional systems often feature enhanced tunability and are more readily accessible in experiments \cite{PhysRevLett.114.043602,Wang:15,PhysRevB.100.075437,PhysRevResearch.4.013185,PhysRevResearch.7.L022033}. For example, 2D SSH lattices are known to harbor higher-order topological phases characterized by corner-localized zero modes \cite{PhysRevLett.118.076803, PhysRevX.9.011012, PhysRevB.107.045118}, and studies of stacked SSH bilayers have revealed interlayer-coupled topological states that interpolate between 1D and 2D behaviors.
However, most studies to date have focused on low-winding-number SSH phases. The systematic construction and characterization of high-winding-number phases with
$|W| \gg 2$-which feature more intricate manifolds of zero-energy edge states-remain comparatively unexplored, despite their potential to host exotic quantum phenomena such as highly degenerate edge modes \cite{PhysRevB.109.035114,PhysRevB.110.165145,PhysRevB.106.085109}, fractional charge localization \cite{unknown,PhysRevLett.95.226801,doi:10.1126/science.aah6442}, and enhanced robustness against disorder \cite{PhysRevB.109.035114,PhysRevB.110.165145}.

Multilayered stacking emerges as a promising strategy to engineer high-winding-number
topological states, as interlayer coupling can modulate the band topology of individual
layers and induce collective nontrivial phases that are absent in singlelayer systems.
This approach echoes recent progress in topological heterostructures: for example,
rhombohedral-stacked graphene multilayers, when coupled to a Haldane model layer,
exhibit high Chern number quantum anomalous Hall (QAH) states with $|C|=N+1$ (where $N$
is the number of graphene layers) \cite{zhao2025high}. In these systems, interlayer
hoppings break the degeneracy of higher-order Dirac points in bulk graphene, while
topological proximity effects from Haldane layers open up bulk band gaps while
preserving nontrivial boundary transport, illustrating how stacking-induced
hybridization gives rise to Chern numbers beyond the singlelayer limit \cite{zhao2025high,sha2024observation,han2024large,guan2024topological,liu2025layer}.

The rhombohedral-stacked structure, with its unique interlayer coupling, is an ideal platform for probing the dependence of topological properties on layer number. The linear amplification of topological invariants with layer count, observed in graphene multilayer systems, provides a novel design strategy, yet this important layer-topology relationship has not been systematically investigated in the SSH model. As a paradigm model of one-dimensional topological systems, it is noted that the topological characteristics of the SSH model are characterized by the winding number $W$ with $W\pm 1$ or $W=0$ corresponding to nontrivial or trivial topological phases. When the SSH model is extended
to a multilayer structure, interlayer transitions will break the topological constraints
of the singlelayer system and induce a richer range of winding number values \cite{LEE202296}.
A key question arises: does the winding number of a rhombohedral-stacked SSH multilayer scale strictly with the number of layers $N$, analogous to the layer-dependent Chern numbers in graphene multilayers?

Regarding the topological characteristics of the SSH model, many experiments have
been carried out to conduct experimental research on it. The theoretical prediction
of the SSH model was verified by measuring the Zak phase \cite{PhysRevLett.134.136601,PhysRevX.4.021017,PhysRevLett.127.147401,PhysRevB.93.041415,PhysRevResearch.4.013185,luo2019topological,2013Direct,xiao2015geometric,PhysRevA.98.032323}, acoustic signal \cite{PhysRevX.4.021017,xiao2015geometric},
optical signal \cite{2013Direct,PhysRevLett.127.147401,PhysRevB.93.041415,PhysRevResearch.4.013185} and electrical signal \cite{luo2019topological,PhysRevA.98.032323}. With the development of quantum
chromatography technology, the Wigner phase space method has gradually become an
alternative approach for distinguishing different quantum states or different
quantum phases. Topological phase transitions are also a type of quantum phase transition. As topological phase transitions constitute a class of quantum phase transitions, and according to the bulk-boundary correspondence, nontrivial topology is manifested in boundary states.
The Wigner phase space method is precisely
based on the quasi-probability distribution of quantum states in the phase space.
From this perspective, the Wigner phase space method seems to be a feasible approach
for detecting topological boundary states.
Many mathematical methods can be used to convert the experimental data from optical homodyne tomography into the state's Wigner function \cite{RevModPhys.81.299}, which makes it possible to experimentally measure the Wigner function directly \cite{douce2013direct,PhysRevLett.78.2547,PhysRevA.60.674}.
Therefore, we will explore whether the
Wigner phase space method can be used to find topologically nontrivial edge states
from a series of quantum states.

\section{The rhombohedral-stacked SSH multilayer system}

\begin{figure}[htp]
		\centering
		\includegraphics[width=0.5\textwidth]{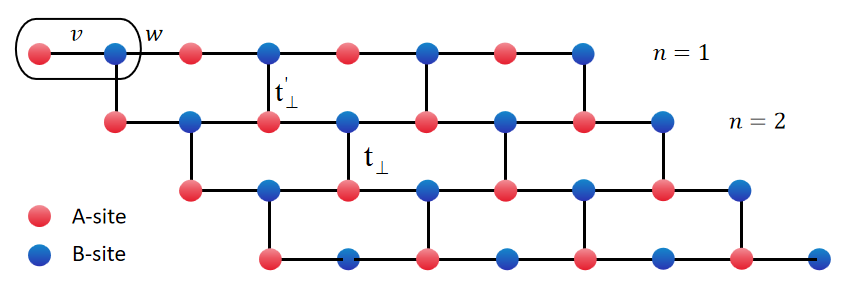}
		\caption{(Color online)
Schematic illustration of the rhombohedral-stacked SSH multilayer system. Each unit cell (indicated by rounded rectangles) contains two sublattice sites: $A$ (red) and $B$ (blue). Within each SSH layer, the intra-cell and inter-cell hopping strengths are denoted by
$\nu$ and $\omega$, respectively. The interlayer coupling is characterized by two distinct hopping parameters:
$t'_{\perp}$ connects sublattice sites vertically between adjacent layers, while
$t_{\perp}$ provides the skewed interlayer connections that establish the rhombohedral stacking geometry.}
		\label{f1}
\end{figure}

The high winding number zero-energy topological states in this paper
are engineered by spinless and non-interacting particles in an
$N$-layer stacked system, which contains $N$ SSH layers
rhombohedral-stacked on each other, as shown in Fig.~\ref{f1}.
The total Hamiltonian of this system is given by
\begin{equation}
\hat{H}=\sum^{N}_{n=1}\hat{H}_{n}+\sum^{N-1}_{n=2}\hat{H}^{\perp}_{n,n+1}+\hat{H}^{\perp}_{H}.
\end{equation}
Each layer denotes the SSH Hamiltonian $\hat{H}_{n}$
on a two-component superlattice \cite{PhysRevLett.42.1698}, and it reads
\begin{equation}
\hat{H}_{n}=\sum_{j,n}\nu c^{\dag}_{j,A,n}c_{j,B,n}+\omega c^{\dag}_{j,B,n}c_{j+1,B,n}+\text{H.c.},
\end{equation}
where $j$ is the super-cell index with $A$ and $B$ the sublattice sites, $\nu$ is the
hopping strength within each super-cell, and $\omega$ is the hopping strength between
nearest-neighbor super-cells. In experiments, the SSH model has been realized in the
cold-atom optical lattice system \cite{2013Direct,PhysRevB.107.054105}, acoustic system \cite{xiao2015geometric,PhysRevX.4.021017}, photonic crystal \cite{PhysRevB.107.165414},
and superconducting quantum circuit \cite{PhysRevA.98.032323}.

The Hamiltonian $H^{\perp}_{n,n+1}$ describing the interlayer
hopping ($n\geq 2$) is given by
\begin{equation}
\hat{H}^{\perp}_{n,n+1}=t_{\perp}\sum_{j}c^{\dag}_{j,B,n}c_{j,A,n+1}+\text{H.c.},
\end{equation}
where $t_{\perp}$ is the interlayer hopping strength. We consider the situation
where the coupling between the first SSH layer and the remaining $N-1$ layer SSH
network structure is weak and freely modulated ($t^{'}_{\perp}\ll t_{\perp}$), whose Hamiltonian $H^{\perp}_{H}$ is given by
\begin{equation}
H^{\perp}_{H}=t^{'}_{\perp}\sum_{j}c^{\dag}_{j,B,0}c_{j,A,1}+\text{H.c.}.
\end{equation}
When $t^{'}_{\perp}=0$, the 1D SSH model can be decoupled from this network structure.

Upon applying the Fourier transformation, we express the Hamiltonian in momentum space as
\begin{equation}
\hat{H}=\sum_{k}\psi^{\dag}_{k}\mathcal{H}_{k}\psi_{k},
\end{equation}
where $\psi_{k}$ represents the annihilation operator arranged as a column vector:
$$\psi_{k}=\left(c_{k,A,0}~c_{k,B,0} \cdots ~c_{k,A,N}~c_{k,B,N} \right)^{T}.$$ 
Here, the subscripts denote momentum $k$, sublattice indices $A$ and $B$, and layer number ranging from 0 to $N$.
The corresponding Bloch Hamiltonian $\mathcal{H}_{k}$ is a $2N\times 2N$ tridiagonal matrix:
\begin{equation}\label{Hk}
\mathcal{H}_{k}=\left(
\begin{array}{cccccccc}
  0 & f_{k} & 0 & 0 & 0 & \cdots & 0 & 0 \\
  f^{*}_{k} & 0 & t^{'}_{\perp} & 0 & 0 & \cdots & 0 & 0 \\
  0 & t^{'}_{\perp} & 0 & f_{k} & 0 & \cdots & 0 & 0 \\
  0 & 0 & f^{*}_{k} & 0 & t_{\perp} & \cdots & 0 & 0 \\
  0 & 0 & 0 & t_{\perp} & 0 & \cdots & 0 & 0 \\
  \vdots & \vdots & \vdots & \vdots & \vdots & \ddots & \vdots & \vdots \\
  0 & 0 & 0 & 0 & 0 & \cdots & 0 & f_{k} \\
  0 & 0 & 0 & 0 & 0 & \cdots & f^{*}_{k} & 0
\end{array}
\right),
\end{equation}
where $f_{k}=\nu+\omega e^{-ik}$.

\section{energy spectrum and winding number}

\begin{figure}[htp]
		\centering
		\includegraphics[width=0.5\textwidth]{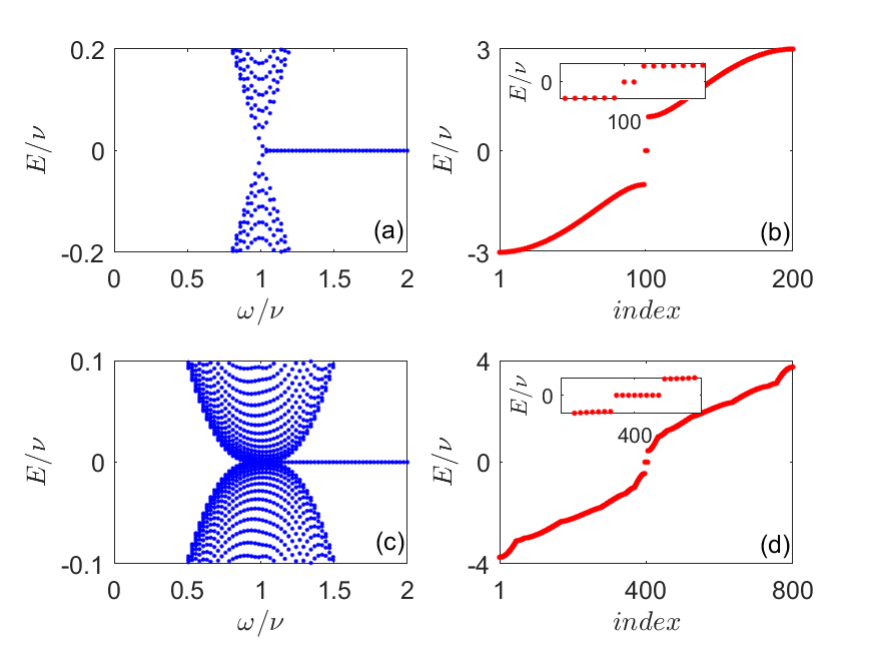}
		\caption{(Color online)
		(a) Band structure under $N=1$.
        (b) Energy versus energy level under $N=1$ and $\omega=2\nu$.
        (c) Band structure under $N=5$.
        (d) Energy versus energy level under $N=5$ and $\omega=2\nu$.
        Other involved parameters are $t'_{\perp}=0.1\nu$, $t_{\perp}=\nu$,
        and $L=100$.
		}
		\label{f2}
\end{figure}

The energy spectrum and eigenstates of the the Hamiltonian in Eq.~(\ref{Hk})
can be obtained numerically via  Schimidt orthogonalization.
We firstly study the decoupling case with $t'_{\perp}=0$. Under $L=100$,
the energy spectrum (lower and higher parts are not shown) as function
of $\omega$ for the 1D SSH model ($N=1$) is plotted in Fig.~\ref{f2}(a).
The zero-energy states are readily seen when $\omega>\nu$. Figure \ref{f2}(b)
shows the energy as a function of the energy
level index for the $\omega=2\nu$ case. It is seen from the inset that
the zero-energy states have $2$-fold degeneracy. Then we study the coupling
case with $t'_{\perp}=0.1\nu$. Taking $N=4$, the low-energy spectrum versus
$\omega$ is plotted in Fig.~\ref{f2}(c). We can see that the zero-energy states
exist when $\omega>\nu$ as well. Although these also are zero-energy states,
we find that the degeneracy of the zero-energy state is closely related to the
number of layers, presenting a $2N$-fold degeneracy feature.

According to the bulk-edge correspondence \cite{PhysRevLett.71.3697}, the winding number $W$
corresponding to the $2$-fold degenerate zero-energy states is $W=1$. Based
on this, it can be inferred that that the winding number corresponding to
the $2N$-fold degenerate zero-energy state shall be $W=N$. For $2N$-band
Bloch systems, it is generally difficult to obtain the analytical solutions.
But we can adopt effective methods to verify the numerical results. Employing the
low-energy theory \cite{Koshino2009}, the effective Hamiltonian of the rhombohedral-stacked $(N-1)$-layer
SSH structure under $\left(c_{k,A,1}~c_{k,B,N}\right)$ is given by
\begin{equation}
H^{SSH}_{N-1}(k)=\left(
\begin{array}{cc}
  0 & F_{N-1}(k) \\
  F^{*}_{N-1}(k) & 0
\end{array}
\right),
\end{equation}
with $F_{N-1}(k)=(-t_{\perp})^{2-N}f^{N}_{k}$, which is gapped system. When the first
SSH layer is stacked on, the total effective Hamiltonian is given by
\begin{equation}
H_{\rm eff}(k)=\left(
\begin{array}{cccc}
  0 & f_{k} & 0 & 0 \\
  f^{*}_{k} & 0 & t'_{\perp} & 0 \\
  0 & t'_{\perp} & 0 & F_{N-1}(k) \\
  0 & 0 & F^{*}_{N-1}(k) & 0
\end{array}
\right).
\end{equation}

When an infinitesimal $t'_{\perp}$ is turned on, the zero-energy states of
the $(N-1)$-layer SSH system still locate in the band gap of the first-layer SSH model.
As mentioned before, it is a fact that the singlelayer SSH model has $2$-fold
degenerate zero-energy states, leading to winding number $W=1$. If the $(N-1)$ SSH multilayers
system has winding number $W=N-1$, the total winding number shall be $W=N$.
Therefore, we transform the problem of proving that the winding number of this $N$-layer
stacked SSH system is $W=N$ into proving that the winding number of the stacked $(N-1)$-layer
SSH system is $W=N-1$. Because an infinitesimal $t'_{\perp}$ is introduced, the
first layer SSH system can be regarded as no affected by the $(N-1)$ SSH multilayers. The
low-energy subspace around the zero energy can by spanned by the basis $(c_{k,A,1}~c_{k,B,N})$.
The low-energy effective can be extracted from $H_{\rm}(k)$ by the Schrieffer-Wolff
transformation as
\begin{equation}
\begin{aligned}
&H^{\rm eff}_{N-1}(k)=H^{SSH}_{N-1}-SH^{-1}_{SSH}S^{\dag} \\
&=
\begin{pmatrix}
  0 & F_{N-1}(k) \\
  F^{*}_{N-1}(k) & 0
\end{pmatrix}
-
\begin{pmatrix}
  0 & 0 \\
  t'_{\perp} & 0
\end{pmatrix}
{\begin{pmatrix}
  0 & f_{k} \\
  f^{*}_{k} & 0
\end{pmatrix}}^{-1}
\begin{pmatrix}
  0 & t'_{\perp} \\
  0 & 0
\end{pmatrix} \\
&=\begin{pmatrix}
0 & F_{N-1}(k) \\
F^{*}_{N-1}(k) & 0
\end{pmatrix},
\end{aligned}
\end{equation}
where $S$ and $S^{\dag}$ is the transformation operators, and $H_{SSH}$ is
the Hamiltonian of the singlelayer SSH model in momentum space.

We then analytically calculate the winding number.
The Hamiltonian $H^{\rm eff}_{N-1}(k)$ scales as
\begin{equation}
\begin{aligned}
H^{\rm eff}_{N-1}(k)& \propto \mathcal{H}^{\rm eff}_{N-1}(k) \\
&=
\begin{pmatrix}
  0 & e^{-i(N-1)\phi(k)} \\
  e^{i(N-1)\phi(k)} & 0
\end{pmatrix},
\end{aligned}
\end{equation}
where $\phi(k)={\rm arctan}\left(\frac{-\sin{k}}{\frac{\nu}{\omega}+\cos{k}}\right)$.
The wave function $|u_{\pm}(k)\rangle$ of $\mathcal{H}^{\rm eff}_{N-1}(k)$ is
given by
\begin{equation}
|u_{\pm}(k)\rangle =\frac{1}{\sqrt{2}}
\begin{pmatrix}
e^{-i(N-1)\phi} \\
\pm 1
\end{pmatrix},
\end{equation}
where $+$ and $-$ corresponds to the positive and negative bands of
$\mathcal{H}^{\rm eff}_{N-1}(k)$, respectively.
With $|u_{-}(k)\rangle$, the Zak phase \cite{PhysRevLett.62.2747} is given by
\begin{equation}
\begin{aligned}
\phi_{\rm Zak}&=i\oint_{-\pi}^{\pi} dk \langle u_{-}(k)| \partial_{k} |u_{-}(k) \rangle \\
&= \frac{1}{2}\oint_{-\pi}^{\pi} dk \frac{\partial\phi(k)}{\partial k} \\
&=
\begin{cases}
0,~~~~~~~~~~~~~~~~~~~~~\omega < \nu \\
(N-1)\pi,~~~~~~~~~~\omega > \nu
\end{cases},
\end{aligned}
\end{equation}
from which we can extract the Winding number $W_{N-1}=\phi_{\rm Zak}/\pi=N-1$.
Together with the first SSH layer, the whole winding number of the stacked $N$-layer
SSH system is $W=W_{N-1}+1\equiv N$. This result is self-consistent with the $2N$-fold
degenerate zero energy states. It indicates that by designing a multilayer rhombohedral-stacked
structure, high winding number zero-energy states can be achieved.

\section{Wigner distribution and Wigner entropy}
Zero-energy states have been detected via Zak phase\cite{2013Direct,PhysRevX.4.021017,xiao2015geometric,PhysRevLett.127.147401,PhysRevB.93.041415,PhysRevResearch.4.013185,luo2019topological,PhysRevA.98.032323},
acoustic \cite{PhysRevX.4.021017,xiao2015geometric},
electrical  \cite{luo2019topological,PhysRevA.98.032323}
and optical signals\cite{2013Direct,PhysRevLett.127.147401,PhysRevB.93.041415,PhysRevResearch.4.013185}.
We now propose an alternative detection method based on the Wigner distribution.
The Wigner function $W(x,p)$ in phase space for a wave function $|\psi\rangle$ is defined as \cite{RevModPhys.81.299,douce2013direct,PhysRevLett.78.2547,PhysRevA.60.674}
\begin{equation}\label{Wigner}
W(x,p)=\frac{1}{2\pi\hbar}\int^{\infty}_{-\infty}\langle{x-\frac{y}{2}}|\rho|{x+\frac{y}{2}}\rangle e^{-\frac{ipy}{\hbar}}dy,
\end{equation}
where $x$ and $p$ denote the coordinate and momentum, respectively, and $\hbar$ is the reduced Planck constant,
and $\rho=|\psi\rangle\langle\psi|$.
\begin{figure}[htp]
	\centering
	\includegraphics[width=0.5\textwidth]{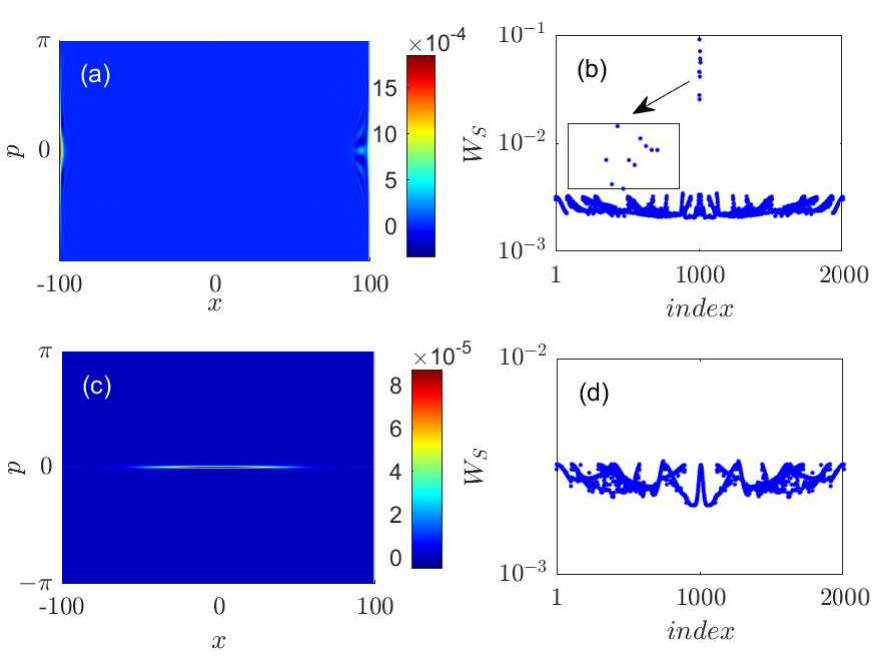}
	\caption{(Color online) 
Distinct signatures of topological edge states versus bulk states in Wigner phase space. (a) Wigner distribution $W (x, p)$ for the 1000-th eigenstate in the topological phase ($\omega=1.5 \nu$), showing pronounced spatial localization at the boundaries along the $x$-axis. (b) Corresponding Wigner entropy $W_s$ as a function of energy level index, with significantly enhanced values for in-gap states (highlighted in inset), providing a clear phase-space diagnostic for topological edge states. (c), (d) Analogous results for the trivial phase ($\omega=0.5 \nu$), where the Wigner distribution remains delocalized throughout the bulk (c) and the Wigner entropy maintains consistently low values across all states (d). The other parameters used are $t_{\perp}^{\prime}=0.1 \nu, t_{\perp}=\nu, N=5$, and $L=200$.
	}
	\label{f3}
\end{figure}
To analyze the boundary localization of zero-energy states, we preprocess the wave function by mapping the probability amplitudes from $N$ layers to a single effective layer:
\begin{equation}
\tilde{|\psi\rangle}=\left[\phi_{1}~\phi_{2}~\cdots~\phi_{L-1}~\phi_{L}\right]^{T},
\end{equation}
where the probability amplitude at site $j$ is
\begin{equation}
\phi_{j}=\sum^{N}_{n=1}\frac{1}{N}\left[\varphi^{2}_{j,A,n}+\varphi^{2}_{j,B,n}\right],
\end{equation}
with $\varphi_{j,A,n}$ and $\varphi_{j,B,n}$ being the original amplitudes. Normalization is not strictly necessary here.
Substituting the preprocessed density matrix $\rho=|\tilde{\psi}\rangle\langle\tilde{\psi}|$ into
Eq.~(\ref{Wigner}) yields the Wigner distribution.

For parameters $t'_{\perp}=0.1\nu$, $t_{\perp}=\nu$, $N=5$, and $L=200$, the Wigner distributions for the
$1000$-th eigenstate at $\omega=1.5\nu$ (topological) and $\omega=0.5\nu$ (trivial) are shown in
Figs.~\ref{f3}(a) and \ref{f3}(c) respectively.
In the topological phase (Figs.~\ref{f3}(a)), $W(x,p)$ exhibits boundary localization along the $x$-axis and a Gaussian distribution along the $p$-axis,characteristic of an edge state. In the trivial phase (Figs.~\ref{f3}(c)), the distribution is Gaussian in $x$ and discrete in $p$, indicating a bulk state.
Manually inspecting Wigner distributions for each state is impractical. We therefore introduce a global quantity, the Wigner entropy, to distinguish edge states from bulk states, and it is defined as
\begin{equation}
W_{s}=-\int\int W(x,p)\ln |W(x,p)|dxdp,
\end{equation}
it quantifies the delocalization in phase space. Fig.~\ref{f3}(b) and Fig.~\ref{f3}(d) plot $W_{S}$versus energy level index for $\omega=1.5\nu$and $\omega=0.5\nu$, respectively.
The Wigner entropy of the zero-energy edge states is significantly larger than that of bulk states, providing a clear signature for identifying topological boundary states from the eigenstate spectrum.

\section{Summary}

We have demonstrated that rhombohedral-stacked $N$-layer SSH systems host $2N$-fold degenerate zero-energy edge states with winding number $W=N$. Through effective Hamiltonian analysis, we proved that weak interlayer coupling (${t}^{\prime}_\perp \ll {t}_\perp$) between the first SSH layer and the remaining (${N}-1$)-layer network preserves the topological properties while enabling linear scaling of the winding number with layer number. The Zak phase calculation confirms $W=N$ in the topologically nontrivial regime ($\omega>\nu$). The introduction of Wigner entropy as a diagnostic tool offers a new phase-space perspective for identifying topological boundary states. The enhanced Wigner entropy of edge states compared to bulk states provides a clear signature that complements traditional detection methods. This phase space approach may prove particularly valuable for experimental characterization of topological states in systems where conventional measurements are challenging.

Our key findings are: (i) The winding number scales exactly as $W=N$, establishing a one-to-one correspondence between layer number and topological invariant; (ii) The $2N$-fold degeneracy of zero-energy states is protected by the rhombohedral stacking geometry; (iii) The Wigner entropy provides a robust experimental signature for detecting topological edge states, with boundary states showing markedly enhanced entropy compared to bulk states in phase space.
These results extend the layer-dependent topological amplification observed in graphene multilayers to one-dimensional systems, demonstrating that rhombohedral stacking is a universal strategy for engineering high-winding-number phases. The highly degenerate edge states may serve as a platform for topological quantum computation and the exploration of fractional charge localization in higher-winding-number regimes.

\newpage

\begin{acknowledgments}
This work is supported by the Natural Science Foundation of Zhejiang Province (Grants No. LQN25A040012), the start-up fund from Xingzhi College, Zhejiang Normal University, and the National Natural Science Foundation of China (Grants No. 12174346).
\end{acknowledgments}

\bibliography{references}

\begin{thebibliography}{47}%
\makeatletter
\providecommand \@ifxundefined [1]{%
 \@ifx{#1\undefined}
}%
\providecommand \@ifnum [1]{%
 \ifnum #1\expandafter \@firstoftwo
 \else \expandafter \@secondoftwo
 \fi
}%
\providecommand \@ifx [1]{%
 \ifx #1\expandafter \@firstoftwo
 \else \expandafter \@secondoftwo
 \fi
}%
\providecommand \natexlab [1]{#1}%
\providecommand \enquote  [1]{``#1''}%
\providecommand \bibnamefont  [1]{#1}%
\providecommand \bibfnamefont [1]{#1}%
\providecommand \citenamefont [1]{#1}%
\providecommand \href@noop [0]{\@secondoftwo}%
\providecommand \href [0]{\begingroup \@sanitize@url \@href}%
\providecommand \@href[1]{\@@startlink{#1}\@@href}%
\providecommand \@@href[1]{\endgroup#1\@@endlink}%
\providecommand \@sanitize@url [0]{\catcode `\\12\catcode `\$12\catcode
  `\&12\catcode `\#12\catcode `\^12\catcode `\_12\catcode `\%12\relax}%
\providecommand \@@startlink[1]{}%
\providecommand \@@endlink[0]{}%
\providecommand \url  [0]{\begingroup\@sanitize@url \@url }%
\providecommand \@url [1]{\endgroup\@href {#1}{\urlprefix }}%
\providecommand \urlprefix  [0]{URL }%
\providecommand \Eprint [0]{\href }%
\providecommand \doibase [0]{https://doi.org/}%
\providecommand \selectlanguage [0]{\@gobble}%
\providecommand \bibinfo  [0]{\@secondoftwo}%
\providecommand \bibfield  [0]{\@secondoftwo}%
\providecommand \translation [1]{[#1]}%
\providecommand \BibitemOpen [0]{}%
\providecommand \bibitemStop [0]{}%
\providecommand \bibitemNoStop [0]{.\EOS\space}%
\providecommand \EOS [0]{\spacefactor3000\relax}%
\providecommand \BibitemShut  [1]{\csname bibitem#1\endcsname}%
\let\auto@bib@innerbib\@empty
\bibitem [{\citenamefont {Hasan}\ and\ \citenamefont
  {Kane}(2010)}]{RevModPhys.82.3045}%
  \BibitemOpen
  \bibfield  {author} {\bibinfo {author} {\bibfnamefont {M.~Z.}\ \bibnamefont
  {Hasan}}\ and\ \bibinfo {author} {\bibfnamefont {C.~L.}\ \bibnamefont
  {Kane}},\ }\bibfield  {title} {\bibinfo {title} {Colloquium: Topological
  insulators},\ }\href {https://doi.org/10.1103/RevModPhys.82.3045} {\bibfield
  {journal} {\bibinfo  {journal} {Rev. Mod. Phys.}\ }\textbf {\bibinfo {volume}
  {82}},\ \bibinfo {pages} {3045} (\bibinfo {year} {2010})}\BibitemShut
  {NoStop}%
\bibitem [{\citenamefont {Shen}(2017)}]{shen2017topological}%
  \BibitemOpen
  \bibfield  {author} {\bibinfo {author} {\bibfnamefont {S.-Q.}\ \bibnamefont
  {Shen}},\ }\href@noop {} {\emph {\bibinfo {title} {Topological Insulators:
  Dirac Equation in Condensed Matters}}},\ \bibinfo {edition} {2nd}\ ed.\
  (\bibinfo  {publisher} {Springer},\ \bibinfo {address} {Singapore},\ \bibinfo
  {year} {2017})\ p.\ \bibinfo {pages} {282}\BibitemShut {NoStop}%
\bibitem [{\citenamefont {Liu}\ \emph {et~al.}(2016)\citenamefont {Liu},
  \citenamefont {Zhang},\ and\ \citenamefont {Qi}}]{liu2016quantum}%
  \BibitemOpen
  \bibfield  {author} {\bibinfo {author} {\bibfnamefont {C.-X.}\ \bibnamefont
  {Liu}}, \bibinfo {author} {\bibfnamefont {S.-C.}\ \bibnamefont {Zhang}},\
  and\ \bibinfo {author} {\bibfnamefont {X.-L.}\ \bibnamefont {Qi}},\
  }\bibfield  {title} {\bibinfo {title} {The quantum anomalous hall effect:
  Theory and experiment},\ }\href@noop {} {\bibfield  {journal} {\bibinfo
  {journal} {Annual Review of Condensed Matter Physics}\ }\textbf {\bibinfo
  {volume} {7}},\ \bibinfo {pages} {301} (\bibinfo {year} {2016})}\BibitemShut
  {NoStop}%
\bibitem [{\citenamefont {Chang}\ \emph {et~al.}(2023)\citenamefont {Chang},
  \citenamefont {Liu},\ and\ \citenamefont {MacDonald}}]{chang2023colloquium}%
  \BibitemOpen
  \bibfield  {author} {\bibinfo {author} {\bibfnamefont {C.-Z.}\ \bibnamefont
  {Chang}}, \bibinfo {author} {\bibfnamefont {C.-X.}\ \bibnamefont {Liu}},\
  and\ \bibinfo {author} {\bibfnamefont {A.~H.}\ \bibnamefont {MacDonald}},\
  }\bibfield  {title} {\bibinfo {title} {Colloquium: Quantum anomalous hall
  effect},\ }\href@noop {} {\bibfield  {journal} {\bibinfo  {journal} {Reviews
  of Modern Physics}\ }\textbf {\bibinfo {volume} {95}},\ \bibinfo {pages}
  {011002} (\bibinfo {year} {2023})}\BibitemShut {NoStop}%
\bibitem [{\citenamefont {Thouless}\ \emph {et~al.}(1982)\citenamefont
  {Thouless}, \citenamefont {Kohmoto}, \citenamefont {Nightingale},\ and\
  \citenamefont {den Nijs}}]{thouless1982quantized}%
  \BibitemOpen
  \bibfield  {author} {\bibinfo {author} {\bibfnamefont {D.~J.}\ \bibnamefont
  {Thouless}}, \bibinfo {author} {\bibfnamefont {M.}~\bibnamefont {Kohmoto}},
  \bibinfo {author} {\bibfnamefont {M.~P.}\ \bibnamefont {Nightingale}},\ and\
  \bibinfo {author} {\bibfnamefont {M.}~\bibnamefont {den Nijs}},\ }\bibfield
  {title} {\bibinfo {title} {Quantized hall conductance in a two-dimensional
  periodic potential},\ }\href@noop {} {\bibfield  {journal} {\bibinfo
  {journal} {Physical Review Letters}\ }\textbf {\bibinfo {volume} {49}},\
  \bibinfo {pages} {405} (\bibinfo {year} {1982})}\BibitemShut {NoStop}%
\bibitem [{\citenamefont {Su}\ \emph {et~al.}(1979{\natexlab{a}})\citenamefont
  {Su}, \citenamefont {Schrieffer},\ and\ \citenamefont
  {Heeger}}]{su1979solitons}%
  \BibitemOpen
  \bibfield  {author} {\bibinfo {author} {\bibfnamefont {W.-P.}\ \bibnamefont
  {Su}}, \bibinfo {author} {\bibfnamefont {J.~R.}\ \bibnamefont {Schrieffer}},\
  and\ \bibinfo {author} {\bibfnamefont {A.~J.}\ \bibnamefont {Heeger}},\
  }\bibfield  {title} {\bibinfo {title} {Solitons in polyacetylene},\
  }\href@noop {} {\bibfield  {journal} {\bibinfo  {journal} {Physical Review
  Letters}\ }\textbf {\bibinfo {volume} {42}},\ \bibinfo {pages} {1698}
  (\bibinfo {year} {1979}{\natexlab{a}})}\BibitemShut {NoStop}%
\bibitem [{\citenamefont {Kivelson}\ \emph {et~al.}(1987)\citenamefont
  {Kivelson}, \citenamefont {Heeger},\ and\ \citenamefont
  {Schrieffer}}]{kivelson1987solitons}%
  \BibitemOpen
  \bibfield  {author} {\bibinfo {author} {\bibfnamefont {S.~A.}\ \bibnamefont
  {Kivelson}}, \bibinfo {author} {\bibfnamefont {A.~J.}\ \bibnamefont
  {Heeger}},\ and\ \bibinfo {author} {\bibfnamefont {J.~R.}\ \bibnamefont
  {Schrieffer}},\ }\bibfield  {title} {\bibinfo {title} {Solitons, polarons,
  and bipolarons in conducting polymers},\ }\href@noop {} {\bibfield  {journal}
  {\bibinfo  {journal} {Reviews of Modern Physics}\ }\textbf {\bibinfo {volume}
  {59}},\ \bibinfo {pages} {845} (\bibinfo {year} {1987})}\BibitemShut
  {NoStop}%
\bibitem [{\citenamefont {Rosenberg}\ and\ \citenamefont
  {Manousakis}(2022)}]{PhysRevB.106.054511}%
  \BibitemOpen
  \bibfield  {author} {\bibinfo {author} {\bibfnamefont {P.}~\bibnamefont
  {Rosenberg}}\ and\ \bibinfo {author} {\bibfnamefont {E.}~\bibnamefont
  {Manousakis}},\ }\bibfield  {title} {\bibinfo {title} {Topological
  superconductivity in a two-dimensional weyl ssh model},\ }\href
  {https://doi.org/10.1103/PhysRevB.106.054511} {\bibfield  {journal} {\bibinfo
   {journal} {Phys. Rev. B}\ }\textbf {\bibinfo {volume} {106}},\ \bibinfo
  {pages} {054511} (\bibinfo {year} {2022})}\BibitemShut {NoStop}%
\bibitem [{\citenamefont {Liu}(2023)}]{PhysRevB.108.245140}%
  \BibitemOpen
  \bibfield  {author} {\bibinfo {author} {\bibfnamefont {F.}~\bibnamefont
  {Liu}},\ }\bibfield  {title} {\bibinfo {title} {Analytic solution of the
  $n$-dimensional su-schrieffer-heeger model},\ }\href
  {https://doi.org/10.1103/PhysRevB.108.245140} {\bibfield  {journal} {\bibinfo
   {journal} {Phys. Rev. B}\ }\textbf {\bibinfo {volume} {108}},\ \bibinfo
  {pages} {245140} (\bibinfo {year} {2023})}\BibitemShut {NoStop}%
\bibitem [{\citenamefont {Zhou}\ \emph {et~al.}(2023)\citenamefont {Zhou},
  \citenamefont {Pan},\ and\ \citenamefont {Jia}}]{PhysRevB.107.054105}%
  \BibitemOpen
  \bibfield  {author} {\bibinfo {author} {\bibfnamefont {X.}~\bibnamefont
  {Zhou}}, \bibinfo {author} {\bibfnamefont {J.-S.}\ \bibnamefont {Pan}},\ and\
  \bibinfo {author} {\bibfnamefont {S.}~\bibnamefont {Jia}},\ }\bibfield
  {title} {\bibinfo {title} {Exploring interacting topological insulator in the
  extended su-schrieffer-heeger model},\ }\href
  {https://doi.org/10.1103/PhysRevB.107.054105} {\bibfield  {journal} {\bibinfo
   {journal} {Phys. Rev. B}\ }\textbf {\bibinfo {volume} {107}},\ \bibinfo
  {pages} {054105} (\bibinfo {year} {2023})}\BibitemShut {NoStop}%
\bibitem [{\citenamefont {Wang}\ \emph
  {et~al.}(2015{\natexlab{a}})\citenamefont {Wang}, \citenamefont {Liu},
  \citenamefont {Zhu},\ and\ \citenamefont {Scully}}]{PhysRevLett.114.043602}%
  \BibitemOpen
  \bibfield  {author} {\bibinfo {author} {\bibfnamefont {D.-W.}\ \bibnamefont
  {Wang}}, \bibinfo {author} {\bibfnamefont {R.-B.}\ \bibnamefont {Liu}},
  \bibinfo {author} {\bibfnamefont {S.-Y.}\ \bibnamefont {Zhu}},\ and\ \bibinfo
  {author} {\bibfnamefont {M.~O.}\ \bibnamefont {Scully}},\ }\bibfield  {title}
  {\bibinfo {title} {Superradiance lattice},\ }\href
  {https://doi.org/10.1103/PhysRevLett.114.043602} {\bibfield  {journal}
  {\bibinfo  {journal} {Phys. Rev. Lett.}\ }\textbf {\bibinfo {volume} {114}},\
  \bibinfo {pages} {043602} (\bibinfo {year} {2015}{\natexlab{a}})}\BibitemShut
  {NoStop}%
\bibitem [{\citenamefont {Wang}\ \emph
  {et~al.}(2015{\natexlab{b}})\citenamefont {Wang}, \citenamefont {Cai},
  \citenamefont {Yuan}, \citenamefont {Zhu},\ and\ \citenamefont
  {Liu}}]{Wang:15}%
  \BibitemOpen
  \bibfield  {author} {\bibinfo {author} {\bibfnamefont {D.-W.}\ \bibnamefont
  {Wang}}, \bibinfo {author} {\bibfnamefont {H.}~\bibnamefont {Cai}}, \bibinfo
  {author} {\bibfnamefont {L.}~\bibnamefont {Yuan}}, \bibinfo {author}
  {\bibfnamefont {S.-Y.}\ \bibnamefont {Zhu}},\ and\ \bibinfo {author}
  {\bibfnamefont {R.-B.}\ \bibnamefont {Liu}},\ }\bibfield  {title} {\bibinfo
  {title} {Topological phase transitions in superradiance lattices},\ }\href
  {https://doi.org/10.1364/OPTICA.2.000712} {\bibfield  {journal} {\bibinfo
  {journal} {Optica}\ }\textbf {\bibinfo {volume} {2}},\ \bibinfo {pages} {712}
  (\bibinfo {year} {2015}{\natexlab{b}})}\BibitemShut {NoStop}%
\bibitem [{\citenamefont {Obana}\ \emph {et~al.}(2019)\citenamefont {Obana},
  \citenamefont {Liu},\ and\ \citenamefont
  {Wakabayashi}}]{PhysRevB.100.075437}%
  \BibitemOpen
  \bibfield  {author} {\bibinfo {author} {\bibfnamefont {D.}~\bibnamefont
  {Obana}}, \bibinfo {author} {\bibfnamefont {F.}~\bibnamefont {Liu}},\ and\
  \bibinfo {author} {\bibfnamefont {K.}~\bibnamefont {Wakabayashi}},\
  }\bibfield  {title} {\bibinfo {title} {Topological edge states in the
  su-schrieffer-heeger model},\ }\href
  {https://doi.org/10.1103/PhysRevB.100.075437} {\bibfield  {journal} {\bibinfo
   {journal} {Phys. Rev. B}\ }\textbf {\bibinfo {volume} {100}},\ \bibinfo
  {pages} {075437} (\bibinfo {year} {2019})}\BibitemShut {NoStop}%
\bibitem [{\citenamefont {C\'aceres-Aravena}\ \emph {et~al.}(2022)\citenamefont
  {C\'aceres-Aravena}, \citenamefont {Real}, \citenamefont {Guzm\'an-Silva},
  \citenamefont {Amo}, \citenamefont {Foa~Torres},\ and\ \citenamefont
  {Vicencio}}]{PhysRevResearch.4.013185}%
  \BibitemOpen
  \bibfield  {author} {\bibinfo {author} {\bibfnamefont {G.}~\bibnamefont
  {C\'aceres-Aravena}}, \bibinfo {author} {\bibfnamefont {B.}~\bibnamefont
  {Real}}, \bibinfo {author} {\bibfnamefont {D.}~\bibnamefont
  {Guzm\'an-Silva}}, \bibinfo {author} {\bibfnamefont {A.}~\bibnamefont {Amo}},
  \bibinfo {author} {\bibfnamefont {L.~E.~F.}\ \bibnamefont {Foa~Torres}},\
  and\ \bibinfo {author} {\bibfnamefont {R.~A.}\ \bibnamefont {Vicencio}},\
  }\bibfield  {title} {\bibinfo {title} {Experimental observation of edge
  states in ssh-stub photonic lattices},\ }\href
  {https://doi.org/10.1103/PhysRevResearch.4.013185} {\bibfield  {journal}
  {\bibinfo  {journal} {Phys. Rev. Res.}\ }\textbf {\bibinfo {volume} {4}},\
  \bibinfo {pages} {013185} (\bibinfo {year} {2022})}\BibitemShut {NoStop}%
\bibitem [{\citenamefont {Khudaiberdiev}\ \emph {et~al.}(2025)\citenamefont
  {Khudaiberdiev}, \citenamefont {Kvon}, \citenamefont {Ryzhkov}, \citenamefont
  {Kozlov}, \citenamefont {Mikhailov},\ and\ \citenamefont
  {Pimenov}}]{PhysRevResearch.7.L022033}%
  \BibitemOpen
  \bibfield  {author} {\bibinfo {author} {\bibfnamefont {D.~A.}\ \bibnamefont
  {Khudaiberdiev}}, \bibinfo {author} {\bibfnamefont {Z.~D.}\ \bibnamefont
  {Kvon}}, \bibinfo {author} {\bibfnamefont {M.~S.}\ \bibnamefont {Ryzhkov}},
  \bibinfo {author} {\bibfnamefont {D.~A.}\ \bibnamefont {Kozlov}}, \bibinfo
  {author} {\bibfnamefont {N.~N.}\ \bibnamefont {Mikhailov}},\ and\ \bibinfo
  {author} {\bibfnamefont {A.}~\bibnamefont {Pimenov}},\ }\bibfield  {title}
  {\bibinfo {title} {Two-dimensional topological anderson insulator in a
  hgte-based semimetal},\ }\href
  {https://doi.org/10.1103/PhysRevResearch.7.L022033} {\bibfield  {journal}
  {\bibinfo  {journal} {Phys. Rev. Res.}\ }\textbf {\bibinfo {volume} {7}},\
  \bibinfo {pages} {L022033} (\bibinfo {year} {2025})}\BibitemShut {NoStop}%
\bibitem [{\citenamefont {Liu}\ and\ \citenamefont
  {Wakabayashi}(2017)}]{PhysRevLett.118.076803}%
  \BibitemOpen
  \bibfield  {author} {\bibinfo {author} {\bibfnamefont {F.}~\bibnamefont
  {Liu}}\ and\ \bibinfo {author} {\bibfnamefont {K.}~\bibnamefont
  {Wakabayashi}},\ }\bibfield  {title} {\bibinfo {title} {Novel topological
  phase with a zero berry curvature},\ }\href
  {https://doi.org/10.1103/PhysRevLett.118.076803} {\bibfield  {journal}
  {\bibinfo  {journal} {Phys. Rev. Lett.}\ }\textbf {\bibinfo {volume} {118}},\
  \bibinfo {pages} {076803} (\bibinfo {year} {2017})}\BibitemShut {NoStop}%
\bibitem [{\citenamefont {Trifunovic}\ and\ \citenamefont
  {Brouwer}(2019)}]{PhysRevX.9.011012}%
  \BibitemOpen
  \bibfield  {author} {\bibinfo {author} {\bibfnamefont {L.}~\bibnamefont
  {Trifunovic}}\ and\ \bibinfo {author} {\bibfnamefont {P.~W.}\ \bibnamefont
  {Brouwer}},\ }\bibfield  {title} {\bibinfo {title} {Higher-order
  bulk-boundary correspondence for topological crystalline phases},\ }\href
  {https://doi.org/10.1103/PhysRevX.9.011012} {\bibfield  {journal} {\bibinfo
  {journal} {Phys. Rev. X}\ }\textbf {\bibinfo {volume} {9}},\ \bibinfo {pages}
  {011012} (\bibinfo {year} {2019})}\BibitemShut {NoStop}%
\bibitem [{\citenamefont {Luo}\ \emph {et~al.}(2023)\citenamefont {Luo},
  \citenamefont {Pan}, \citenamefont {Liu},\ and\ \citenamefont
  {Liu}}]{PhysRevB.107.045118}%
  \BibitemOpen
  \bibfield  {author} {\bibinfo {author} {\bibfnamefont {X.-J.}\ \bibnamefont
  {Luo}}, \bibinfo {author} {\bibfnamefont {X.-H.}\ \bibnamefont {Pan}},
  \bibinfo {author} {\bibfnamefont {C.-X.}\ \bibnamefont {Liu}},\ and\ \bibinfo
  {author} {\bibfnamefont {X.}~\bibnamefont {Liu}},\ }\bibfield  {title}
  {\bibinfo {title} {Higher-order topological phases emerging from
  su-schrieffer-heeger stacking},\ }\href
  {https://doi.org/10.1103/PhysRevB.107.045118} {\bibfield  {journal} {\bibinfo
   {journal} {Phys. Rev. B}\ }\textbf {\bibinfo {volume} {107}},\ \bibinfo
  {pages} {045118} (\bibinfo {year} {2023})}\BibitemShut {NoStop}%
\bibitem [{\citenamefont {Cinnirella}\ \emph {et~al.}(2024)\citenamefont
  {Cinnirella}, \citenamefont {Nava}, \citenamefont {Campagnano},\ and\
  \citenamefont {Giuliano}}]{PhysRevB.109.035114}%
  \BibitemOpen
  \bibfield  {author} {\bibinfo {author} {\bibfnamefont {E.~G.}\ \bibnamefont
  {Cinnirella}}, \bibinfo {author} {\bibfnamefont {A.}~\bibnamefont {Nava}},
  \bibinfo {author} {\bibfnamefont {G.}~\bibnamefont {Campagnano}},\ and\
  \bibinfo {author} {\bibfnamefont {D.}~\bibnamefont {Giuliano}},\ }\bibfield
  {title} {\bibinfo {title} {Fate of high winding number topological phases in
  the disordered extended su-schrieffer-heeger model},\ }\href
  {https://doi.org/10.1103/PhysRevB.109.035114} {\bibfield  {journal} {\bibinfo
   {journal} {Phys. Rev. B}\ }\textbf {\bibinfo {volume} {109}},\ \bibinfo
  {pages} {035114} (\bibinfo {year} {2024})}\BibitemShut {NoStop}%
\bibitem [{\citenamefont {Di~Salvo}\ \emph {et~al.}(2024)\citenamefont
  {Di~Salvo}, \citenamefont {Moustaj}, \citenamefont {Xu}, \citenamefont
  {Fritz}, \citenamefont {Mitchell}, \citenamefont {Smith},\ and\ \citenamefont
  {Schuricht}}]{PhysRevB.110.165145}%
  \BibitemOpen
  \bibfield  {author} {\bibinfo {author} {\bibfnamefont {E.}~\bibnamefont
  {Di~Salvo}}, \bibinfo {author} {\bibfnamefont {A.}~\bibnamefont {Moustaj}},
  \bibinfo {author} {\bibfnamefont {C.}~\bibnamefont {Xu}}, \bibinfo {author}
  {\bibfnamefont {L.}~\bibnamefont {Fritz}}, \bibinfo {author} {\bibfnamefont
  {A.~K.}\ \bibnamefont {Mitchell}}, \bibinfo {author} {\bibfnamefont {C.~M.}\
  \bibnamefont {Smith}},\ and\ \bibinfo {author} {\bibfnamefont
  {D.}~\bibnamefont {Schuricht}},\ }\bibfield  {title} {\bibinfo {title}
  {Topological phases of the interacting su-schrieffer-heeger model: An
  analytical study},\ }\href {https://doi.org/10.1103/PhysRevB.110.165145}
  {\bibfield  {journal} {\bibinfo  {journal} {Phys. Rev. B}\ }\textbf {\bibinfo
  {volume} {110}},\ \bibinfo {pages} {165145} (\bibinfo {year}
  {2024})}\BibitemShut {NoStop}%
\bibitem [{\citenamefont {Anastasiadis}\ \emph {et~al.}(2022)\citenamefont
  {Anastasiadis}, \citenamefont {Styliaris}, \citenamefont {Chaunsali},
  \citenamefont {Theocharis},\ and\ \citenamefont
  {Diakonos}}]{PhysRevB.106.085109}%
  \BibitemOpen
  \bibfield  {author} {\bibinfo {author} {\bibfnamefont {A.}~\bibnamefont
  {Anastasiadis}}, \bibinfo {author} {\bibfnamefont {G.}~\bibnamefont
  {Styliaris}}, \bibinfo {author} {\bibfnamefont {R.}~\bibnamefont
  {Chaunsali}}, \bibinfo {author} {\bibfnamefont {G.}~\bibnamefont
  {Theocharis}},\ and\ \bibinfo {author} {\bibfnamefont {F.~K.}\ \bibnamefont
  {Diakonos}},\ }\bibfield  {title} {\bibinfo {title} {Bulk-edge correspondence
  in the trimer su-schrieffer-heeger model},\ }\href
  {https://doi.org/10.1103/PhysRevB.106.085109} {\bibfield  {journal} {\bibinfo
   {journal} {Phys. Rev. B}\ }\textbf {\bibinfo {volume} {106}},\ \bibinfo
  {pages} {085109} (\bibinfo {year} {2022})}\BibitemShut {NoStop}%
\bibitem [{\citenamefont {Wu}(2022)}]{unknown}%
  \BibitemOpen
  \bibfield  {author} {\bibinfo {author} {\bibfnamefont {Y.}~\bibnamefont
  {Wu}},\ }\bibfield  {title} {\bibinfo {title} {Fractional charges in the
  su-schrieffer-heeger model},\ }\href
  {https://doi.org/10.48550/arXiv.2209.05981} {\  (\bibinfo {year}
  {2022})}\BibitemShut {NoStop}%
\bibitem [{\citenamefont {Kane}\ and\ \citenamefont
  {Mele}(2005)}]{PhysRevLett.95.226801}%
  \BibitemOpen
  \bibfield  {author} {\bibinfo {author} {\bibfnamefont {C.~L.}\ \bibnamefont
  {Kane}}\ and\ \bibinfo {author} {\bibfnamefont {E.~J.}\ \bibnamefont
  {Mele}},\ }\bibfield  {title} {\bibinfo {title} {Quantum spin hall effect in
  graphene},\ }\href {https://doi.org/10.1103/PhysRevLett.95.226801} {\bibfield
   {journal} {\bibinfo  {journal} {Phys. Rev. Lett.}\ }\textbf {\bibinfo
  {volume} {95}},\ \bibinfo {pages} {226801} (\bibinfo {year}
  {2005})}\BibitemShut {NoStop}%
\bibitem [{\citenamefont {Benalcazar}\ \emph {et~al.}(2017)\citenamefont
  {Benalcazar}, \citenamefont {Bernevig},\ and\ \citenamefont
  {Hughes}}]{doi:10.1126/science.aah6442}%
  \BibitemOpen
  \bibfield  {author} {\bibinfo {author} {\bibfnamefont {W.~A.}\ \bibnamefont
  {Benalcazar}}, \bibinfo {author} {\bibfnamefont {B.~A.}\ \bibnamefont
  {Bernevig}},\ and\ \bibinfo {author} {\bibfnamefont {T.~L.}\ \bibnamefont
  {Hughes}},\ }\bibfield  {title} {\bibinfo {title} {Quantized electric
  multipole insulators},\ }\href {https://doi.org/10.1126/science.aah6442}
  {\bibfield  {journal} {\bibinfo  {journal} {Science}\ }\textbf {\bibinfo
  {volume} {357}},\ \bibinfo {pages} {61} (\bibinfo {year} {2017})},\ \Eprint
  {https://arxiv.org/abs/https://www.science.org/doi/pdf/10.1126/science.aah6442}
  {https://www.science.org/doi/pdf/10.1126/science.aah6442} \BibitemShut
  {NoStop}%
\bibitem [{\citenamefont {Zhao}\ \emph {et~al.}(2025)\citenamefont {Zhao},
  \citenamefont {Zhang},\ and\ \citenamefont {Zhang}}]{zhao2025high}%
  \BibitemOpen
  \bibfield  {author} {\bibinfo {author} {\bibfnamefont {Y.}~\bibnamefont
  {Zhao}}, \bibinfo {author} {\bibfnamefont {L.}~\bibnamefont {Zhang}},\ and\
  \bibinfo {author} {\bibfnamefont {F.-C.}\ \bibnamefont {Zhang}},\ }\bibfield
  {title} {\bibinfo {title} {High chern number quantum anomalous hall states in
  haldane-graphene multilayers},\ }\href@noop {} {\bibfield  {journal}
  {\bibinfo  {journal} {[Submitted]}\ } (\bibinfo {year} {2025})},\ \bibinfo
  {note} {document: LE20304B.pdf}\BibitemShut {NoStop}%
\bibitem [{\citenamefont {Sha}\ \emph {et~al.}(2024)\citenamefont {Sha},
  \citenamefont {Han}, \citenamefont {Liu} \emph
  {et~al.}}]{sha2024observation}%
  \BibitemOpen
  \bibfield  {author} {\bibinfo {author} {\bibfnamefont {Y.}~\bibnamefont
  {Sha}}, \bibinfo {author} {\bibfnamefont {T.}~\bibnamefont {Han}}, \bibinfo
  {author} {\bibfnamefont {Z.}~\bibnamefont {Liu}}, \emph {et~al.},\ }\bibfield
   {title} {\bibinfo {title} {Observation of a chern insulator in crystalline
  abca-tetralayer graphene with spin-orbit coupling},\ }\href@noop {}
  {\bibfield  {journal} {\bibinfo  {journal} {Science}\ }\textbf {\bibinfo
  {volume} {384}},\ \bibinfo {pages} {414} (\bibinfo {year}
  {2024})}\BibitemShut {NoStop}%
\bibitem [{\citenamefont {Han}\ \emph {et~al.}(2024)\citenamefont {Han},
  \citenamefont {Lu}, \citenamefont {Yao}, \citenamefont {Yang}, \citenamefont
  {Seo}, \citenamefont {Yoon}, \citenamefont {Watanabe}, \citenamefont
  {Taniguchi}, \citenamefont {Fu}, \citenamefont {Zhang},\ and\ \citenamefont
  {Ju}}]{han2024large}%
  \BibitemOpen
  \bibfield  {author} {\bibinfo {author} {\bibfnamefont {T.}~\bibnamefont
  {Han}}, \bibinfo {author} {\bibfnamefont {Z.}~\bibnamefont {Lu}}, \bibinfo
  {author} {\bibfnamefont {Y.}~\bibnamefont {Yao}}, \bibinfo {author}
  {\bibfnamefont {J.}~\bibnamefont {Yang}}, \bibinfo {author} {\bibfnamefont
  {J.}~\bibnamefont {Seo}}, \bibinfo {author} {\bibfnamefont {C.}~\bibnamefont
  {Yoon}}, \bibinfo {author} {\bibfnamefont {K.}~\bibnamefont {Watanabe}},
  \bibinfo {author} {\bibfnamefont {T.}~\bibnamefont {Taniguchi}}, \bibinfo
  {author} {\bibfnamefont {L.}~\bibnamefont {Fu}}, \bibinfo {author}
  {\bibfnamefont {F.}~\bibnamefont {Zhang}},\ and\ \bibinfo {author}
  {\bibfnamefont {L.}~\bibnamefont {Ju}},\ }\bibfield  {title} {\bibinfo
  {title} {Large quantum anomalous hall effect in spin-orbit proximitized
  rhombohedral graphene},\ }\href {https://doi.org/10.1126/science.adk9749}
  {\bibfield  {journal} {\bibinfo  {journal} {Science}\ }\textbf {\bibinfo
  {volume} {384}},\ \bibinfo {pages} {647} (\bibinfo {year} {2024})},\ \Eprint
  {https://arxiv.org/abs/https://www.science.org/doi/pdf/10.1126/science.adk9749}
  {https://www.science.org/doi/pdf/10.1126/science.adk9749} \BibitemShut
  {NoStop}%
\bibitem [{\citenamefont {Guan}\ \emph {et~al.}(2024)\citenamefont {Guan},
  \citenamefont {Lou},\ and\ \citenamefont {Chang}}]{guan2024topological}%
  \BibitemOpen
  \bibfield  {author} {\bibinfo {author} {\bibfnamefont {J.-H.}\ \bibnamefont
  {Guan}}, \bibinfo {author} {\bibfnamefont {W.-K.}\ \bibnamefont {Lou}},\ and\
  \bibinfo {author} {\bibfnamefont {K.}~\bibnamefont {Chang}},\ }\bibfield
  {title} {\bibinfo {title} {Topological hidden phase transition in honeycomb
  bilayers with a high chern number},\ }\href@noop {} {\bibfield  {journal}
  {\bibinfo  {journal} {Physical Review B}\ }\textbf {\bibinfo {volume}
  {110}},\ \bibinfo {pages} {165303} (\bibinfo {year} {2024})}\BibitemShut
  {NoStop}%
\bibitem [{\citenamefont {Liu}\ and\ \citenamefont
  {Wang}(2025)}]{liu2025layer}%
  \BibitemOpen
  \bibfield  {author} {\bibinfo {author} {\bibfnamefont {Z.}~\bibnamefont
  {Liu}}\ and\ \bibinfo {author} {\bibfnamefont {J.}~\bibnamefont {Wang}},\
  }\bibfield  {title} {\bibinfo {title} {Layer-dependent quantum anomalous hall
  effect in rhombohedral graphene},\ }\href@noop {} {\bibfield  {journal}
  {\bibinfo  {journal} {Physical Review B}\ }\textbf {\bibinfo {volume}
  {111}},\ \bibinfo {pages} {L081111} (\bibinfo {year} {2025})}\BibitemShut
  {NoStop}%
\bibitem [{\citenamefont {Lee}\ \emph {et~al.}(2022)\citenamefont {Lee},
  \citenamefont {Io},\ and\ \citenamefont {chung Kao}}]{LEE202296}%
  \BibitemOpen
  \bibfield  {author} {\bibinfo {author} {\bibfnamefont {C.-S.}\ \bibnamefont
  {Lee}}, \bibinfo {author} {\bibfnamefont {I.-F.}\ \bibnamefont {Io}},\ and\
  \bibinfo {author} {\bibfnamefont {H.}~\bibnamefont {chung Kao}},\ }\bibfield
  {title} {\bibinfo {title} {Winding number and zak phase in multi-band ssh
  models},\ }\href {https://doi.org/https://doi.org/10.1016/j.cjph.2022.05.007}
  {\bibfield  {journal} {\bibinfo  {journal} {Chinese Journal of Physics}\
  }\textbf {\bibinfo {volume} {78}},\ \bibinfo {pages} {96} (\bibinfo {year}
  {2022})}\BibitemShut {NoStop}%
\bibitem [{\citenamefont {Chen}\ \emph {et~al.}(2025)\citenamefont {Chen},
  \citenamefont {Zhang}, \citenamefont {Sun}, \citenamefont {Zhang},
  \citenamefont {Tang}, \citenamefont {Yang}, \citenamefont {Zhu},\ and\
  \citenamefont {Lu}}]{PhysRevLett.134.136601}%
  \BibitemOpen
  \bibfield  {author} {\bibinfo {author} {\bibfnamefont {Z.-X.}\ \bibnamefont
  {Chen}}, \bibinfo {author} {\bibfnamefont {Y.-H.}\ \bibnamefont {Zhang}},
  \bibinfo {author} {\bibfnamefont {X.-C.}\ \bibnamefont {Sun}}, \bibinfo
  {author} {\bibfnamefont {R.-Y.}\ \bibnamefont {Zhang}}, \bibinfo {author}
  {\bibfnamefont {J.-S.}\ \bibnamefont {Tang}}, \bibinfo {author}
  {\bibfnamefont {X.}~\bibnamefont {Yang}}, \bibinfo {author} {\bibfnamefont
  {X.-F.}\ \bibnamefont {Zhu}},\ and\ \bibinfo {author} {\bibfnamefont {Y.-Q.}\
  \bibnamefont {Lu}},\ }\bibfield  {title} {\bibinfo {title} {Direct
  measurement of topological invariants through temporal adiabatic evolution of
  bulk states in the synthetic brillouin zone},\ }\href
  {https://doi.org/10.1103/PhysRevLett.134.136601} {\bibfield  {journal}
  {\bibinfo  {journal} {Phys. Rev. Lett.}\ }\textbf {\bibinfo {volume} {134}},\
  \bibinfo {pages} {136601} (\bibinfo {year} {2025})}\BibitemShut {NoStop}%
\bibitem [{\citenamefont {Xiao}\ \emph {et~al.}(2014)\citenamefont {Xiao},
  \citenamefont {Zhang},\ and\ \citenamefont {Chan}}]{PhysRevX.4.021017}%
  \BibitemOpen
  \bibfield  {author} {\bibinfo {author} {\bibfnamefont {M.}~\bibnamefont
  {Xiao}}, \bibinfo {author} {\bibfnamefont {Z.~Q.}\ \bibnamefont {Zhang}},\
  and\ \bibinfo {author} {\bibfnamefont {C.~T.}\ \bibnamefont {Chan}},\
  }\bibfield  {title} {\bibinfo {title} {Surface impedance and bulk band
  geometric phases in one-dimensional systems},\ }\href
  {https://doi.org/10.1103/PhysRevX.4.021017} {\bibfield  {journal} {\bibinfo
  {journal} {Phys. Rev. X}\ }\textbf {\bibinfo {volume} {4}},\ \bibinfo {pages}
  {021017} (\bibinfo {year} {2014})}\BibitemShut {NoStop}%
\bibitem [{\citenamefont {Jiao}\ \emph {et~al.}(2021)\citenamefont {Jiao},
  \citenamefont {Longhi}, \citenamefont {Wang}, \citenamefont {Gao},
  \citenamefont {Zhou}, \citenamefont {Wang}, \citenamefont {Fu}, \citenamefont
  {Wang}, \citenamefont {Ren}, \citenamefont {Qiao},\ and\ \citenamefont
  {Jin}}]{PhysRevLett.127.147401}%
  \BibitemOpen
  \bibfield  {author} {\bibinfo {author} {\bibfnamefont {Z.-Q.}\ \bibnamefont
  {Jiao}}, \bibinfo {author} {\bibfnamefont {S.}~\bibnamefont {Longhi}},
  \bibinfo {author} {\bibfnamefont {X.-W.}\ \bibnamefont {Wang}}, \bibinfo
  {author} {\bibfnamefont {J.}~\bibnamefont {Gao}}, \bibinfo {author}
  {\bibfnamefont {W.-H.}\ \bibnamefont {Zhou}}, \bibinfo {author}
  {\bibfnamefont {Y.}~\bibnamefont {Wang}}, \bibinfo {author} {\bibfnamefont
  {Y.-X.}\ \bibnamefont {Fu}}, \bibinfo {author} {\bibfnamefont
  {L.}~\bibnamefont {Wang}}, \bibinfo {author} {\bibfnamefont {R.-J.}\
  \bibnamefont {Ren}}, \bibinfo {author} {\bibfnamefont {L.-F.}\ \bibnamefont
  {Qiao}},\ and\ \bibinfo {author} {\bibfnamefont {X.-M.}\ \bibnamefont
  {Jin}},\ }\bibfield  {title} {\bibinfo {title} {Experimentally detecting
  quantized zak phases without chiral symmetry in photonic lattices},\ }\href
  {https://doi.org/10.1103/PhysRevLett.127.147401} {\bibfield  {journal}
  {\bibinfo  {journal} {Phys. Rev. Lett.}\ }\textbf {\bibinfo {volume} {127}},\
  \bibinfo {pages} {147401} (\bibinfo {year} {2021})}\BibitemShut {NoStop}%
\bibitem [{\citenamefont {Wang}\ \emph {et~al.}(2016)\citenamefont {Wang},
  \citenamefont {Xiao}, \citenamefont {Liu}, \citenamefont {Zhu},\ and\
  \citenamefont {Chan}}]{PhysRevB.93.041415}%
  \BibitemOpen
  \bibfield  {author} {\bibinfo {author} {\bibfnamefont {Q.}~\bibnamefont
  {Wang}}, \bibinfo {author} {\bibfnamefont {M.}~\bibnamefont {Xiao}}, \bibinfo
  {author} {\bibfnamefont {H.}~\bibnamefont {Liu}}, \bibinfo {author}
  {\bibfnamefont {S.}~\bibnamefont {Zhu}},\ and\ \bibinfo {author}
  {\bibfnamefont {C.~T.}\ \bibnamefont {Chan}},\ }\bibfield  {title} {\bibinfo
  {title} {Measurement of the zak phase of photonic bands through the interface
  states of a metasurface/photonic crystal},\ }\href
  {https://doi.org/10.1103/PhysRevB.93.041415} {\bibfield  {journal} {\bibinfo
  {journal} {Phys. Rev. B}\ }\textbf {\bibinfo {volume} {93}},\ \bibinfo
  {pages} {041415} (\bibinfo {year} {2016})}\BibitemShut {NoStop}%
\bibitem [{\citenamefont {Luo}\ and\ \citenamefont
  {Yu}(2019)}]{luo2019topological}%
  \BibitemOpen
  \bibfield  {author} {\bibinfo {author} {\bibfnamefont {K.-F.}\ \bibnamefont
  {Luo}}\ and\ \bibinfo {author} {\bibfnamefont {R.}~\bibnamefont {Yu}},\
  }\bibfield  {title} {\bibinfo {title} {Topological states in electric
  circuit},\ }\href {https://doi.org/10.7498/aps.68.20191398} {\bibfield
  {journal} {\bibinfo  {journal} {Acta Phys. Sin.}\ }\textbf {\bibinfo {volume}
  {68}},\ \bibinfo {pages} {220305} (\bibinfo {year} {2019})}\BibitemShut
  {NoStop}%
\bibitem [{\citenamefont {Atala}\ \emph {et~al.}(2013)\citenamefont {Atala},
  \citenamefont {Aidelsburger}, \citenamefont {Barreiro}, \citenamefont
  {Abanin},\ and\ \citenamefont {Bloch}}]{2013Direct}%
  \BibitemOpen
  \bibfield  {author} {\bibinfo {author} {\bibfnamefont {M.}~\bibnamefont
  {Atala}}, \bibinfo {author} {\bibfnamefont {M.}~\bibnamefont {Aidelsburger}},
  \bibinfo {author} {\bibfnamefont {J.~T.}\ \bibnamefont {Barreiro}}, \bibinfo
  {author} {\bibfnamefont {D.}~\bibnamefont {Abanin}},\ and\ \bibinfo {author}
  {\bibfnamefont {I.}~\bibnamefont {Bloch}},\ }\bibfield  {title} {\bibinfo
  {title} {Direct measurement of the zak phase in topological bloch bands},\
  }\href@noop {} {\bibfield  {journal} {\bibinfo  {journal} {Nature Physics}\
  }\textbf {\bibinfo {volume} {9}},\ \bibinfo {pages} {795} (\bibinfo {year}
  {2013})}\BibitemShut {NoStop}%
\bibitem [{\citenamefont {Xiao}\ \emph {et~al.}(2015)\citenamefont {Xiao},
  \citenamefont {Ma}, \citenamefont {Yang}, \citenamefont {Sheng},
  \citenamefont {Zhang},\ and\ \citenamefont {Chan}}]{xiao2015geometric}%
  \BibitemOpen
  \bibfield  {author} {\bibinfo {author} {\bibfnamefont {M.}~\bibnamefont
  {Xiao}}, \bibinfo {author} {\bibfnamefont {G.}~\bibnamefont {Ma}}, \bibinfo
  {author} {\bibfnamefont {Z.}~\bibnamefont {Yang}}, \bibinfo {author}
  {\bibfnamefont {P.}~\bibnamefont {Sheng}}, \bibinfo {author} {\bibfnamefont
  {Z.~Q.}\ \bibnamefont {Zhang}},\ and\ \bibinfo {author} {\bibfnamefont
  {C.~T.}\ \bibnamefont {Chan}},\ }\bibfield  {title} {\bibinfo {title}
  {Geometric phase and band inversion in periodic acoustic systems},\ }\href
  {https://doi.org/10.1038/nphys3228} {\bibfield  {journal} {\bibinfo
  {journal} {Nature Physics}\ }\textbf {\bibinfo {volume} {11}},\ \bibinfo
  {pages} {240} (\bibinfo {year} {2015})}\BibitemShut {NoStop}%
\bibitem [{\citenamefont {Mei}\ \emph {et~al.}(2018)\citenamefont {Mei},
  \citenamefont {Chen}, \citenamefont {Tian}, \citenamefont {Zhu},\ and\
  \citenamefont {Jia}}]{PhysRevA.98.032323}%
  \BibitemOpen
  \bibfield  {author} {\bibinfo {author} {\bibfnamefont {F.}~\bibnamefont
  {Mei}}, \bibinfo {author} {\bibfnamefont {G.}~\bibnamefont {Chen}}, \bibinfo
  {author} {\bibfnamefont {L.}~\bibnamefont {Tian}}, \bibinfo {author}
  {\bibfnamefont {S.-L.}\ \bibnamefont {Zhu}},\ and\ \bibinfo {author}
  {\bibfnamefont {S.}~\bibnamefont {Jia}},\ }\bibfield  {title} {\bibinfo
  {title} {Topology-dependent quantum dynamics and entanglement-dependent
  topological pumping in superconducting qubit chains},\ }\href
  {https://doi.org/10.1103/PhysRevA.98.032323} {\bibfield  {journal} {\bibinfo
  {journal} {Phys. Rev. A}\ }\textbf {\bibinfo {volume} {98}},\ \bibinfo
  {pages} {032323} (\bibinfo {year} {2018})}\BibitemShut {NoStop}%
\bibitem [{\citenamefont {Lvovsky}\ and\ \citenamefont
  {Raymer}(2009)}]{RevModPhys.81.299}%
  \BibitemOpen
  \bibfield  {author} {\bibinfo {author} {\bibfnamefont {A.~I.}\ \bibnamefont
  {Lvovsky}}\ and\ \bibinfo {author} {\bibfnamefont {M.~G.}\ \bibnamefont
  {Raymer}},\ }\bibfield  {title} {\bibinfo {title} {Continuous-variable
  optical quantum-state tomography},\ }\href
  {https://doi.org/10.1103/RevModPhys.81.299} {\bibfield  {journal} {\bibinfo
  {journal} {Rev. Mod. Phys.}\ }\textbf {\bibinfo {volume} {81}},\ \bibinfo
  {pages} {299} (\bibinfo {year} {2009})}\BibitemShut {NoStop}%
\bibitem [{\citenamefont {Douce}\ \emph {et~al.}(2013)\citenamefont {Douce},
  \citenamefont {Eckstein}, \citenamefont {Walborn}, \citenamefont {Khoury},
  \citenamefont {Ducci}, \citenamefont {Keller}, \citenamefont {Coudreau},\
  and\ \citenamefont {Milman}}]{douce2013direct}%
  \BibitemOpen
  \bibfield  {author} {\bibinfo {author} {\bibfnamefont {T.}~\bibnamefont
  {Douce}}, \bibinfo {author} {\bibfnamefont {A.}~\bibnamefont {Eckstein}},
  \bibinfo {author} {\bibfnamefont {S.~P.}\ \bibnamefont {Walborn}}, \bibinfo
  {author} {\bibfnamefont {A.~Z.}\ \bibnamefont {Khoury}}, \bibinfo {author}
  {\bibfnamefont {S.}~\bibnamefont {Ducci}}, \bibinfo {author} {\bibfnamefont
  {A.}~\bibnamefont {Keller}}, \bibinfo {author} {\bibfnamefont
  {T.}~\bibnamefont {Coudreau}},\ and\ \bibinfo {author} {\bibfnamefont
  {P.}~\bibnamefont {Milman}},\ }\bibfield  {title} {\bibinfo {title} {Direct
  measurement of the biphoton wigner function through two-photon
  interference},\ }\href {https://doi.org/10.1038/srep03530} {\bibfield
  {journal} {\bibinfo  {journal} {Sci. Rep.}\ }\textbf {\bibinfo {volume}
  {3}},\ \bibinfo {pages} {3530} (\bibinfo {year} {2013})}\BibitemShut
  {NoStop}%
\bibitem [{\citenamefont {Lutterbach}\ and\ \citenamefont
  {Davidovich}(1997)}]{PhysRevLett.78.2547}%
  \BibitemOpen
  \bibfield  {author} {\bibinfo {author} {\bibfnamefont {L.~G.}\ \bibnamefont
  {Lutterbach}}\ and\ \bibinfo {author} {\bibfnamefont {L.}~\bibnamefont
  {Davidovich}},\ }\bibfield  {title} {\bibinfo {title} {Method for direct
  measurement of the wigner function in cavity qed and ion traps},\ }\href
  {https://doi.org/10.1103/PhysRevLett.78.2547} {\bibfield  {journal} {\bibinfo
   {journal} {Phys. Rev. Lett.}\ }\textbf {\bibinfo {volume} {78}},\ \bibinfo
  {pages} {2547} (\bibinfo {year} {1997})}\BibitemShut {NoStop}%
\bibitem [{\citenamefont {Banaszek}\ \emph {et~al.}(1999)\citenamefont
  {Banaszek}, \citenamefont {Radzewicz}, \citenamefont {W\'odkiewicz},\ and\
  \citenamefont {Krasi\ifmmode~\acute{n}\else
  \'{n}\fi{}ski}}]{PhysRevA.60.674}%
  \BibitemOpen
  \bibfield  {author} {\bibinfo {author} {\bibfnamefont {K.}~\bibnamefont
  {Banaszek}}, \bibinfo {author} {\bibfnamefont {C.}~\bibnamefont {Radzewicz}},
  \bibinfo {author} {\bibfnamefont {K.}~\bibnamefont {W\'odkiewicz}},\ and\
  \bibinfo {author} {\bibfnamefont {J.~S.}\ \bibnamefont
  {Krasi\ifmmode~\acute{n}\else \'{n}\fi{}ski}},\ }\bibfield  {title} {\bibinfo
  {title} {Direct measurement of the wigner function by photon counting},\
  }\href {https://doi.org/10.1103/PhysRevA.60.674} {\bibfield  {journal}
  {\bibinfo  {journal} {Phys. Rev. A}\ }\textbf {\bibinfo {volume} {60}},\
  \bibinfo {pages} {674} (\bibinfo {year} {1999})}\BibitemShut {NoStop}%
\bibitem [{\citenamefont {Su}\ \emph {et~al.}(1979{\natexlab{b}})\citenamefont
  {Su}, \citenamefont {Schrieffer},\ and\ \citenamefont
  {Heeger}}]{PhysRevLett.42.1698}%
  \BibitemOpen
  \bibfield  {author} {\bibinfo {author} {\bibfnamefont {W.~P.}\ \bibnamefont
  {Su}}, \bibinfo {author} {\bibfnamefont {J.~R.}\ \bibnamefont {Schrieffer}},\
  and\ \bibinfo {author} {\bibfnamefont {A.~J.}\ \bibnamefont {Heeger}},\
  }\bibfield  {title} {\bibinfo {title} {Solitons in polyacetylene},\ }\href
  {https://doi.org/10.1103/PhysRevLett.42.1698} {\bibfield  {journal} {\bibinfo
   {journal} {Phys. Rev. Lett.}\ }\textbf {\bibinfo {volume} {42}},\ \bibinfo
  {pages} {1698} (\bibinfo {year} {1979}{\natexlab{b}})}\BibitemShut {NoStop}%
\bibitem [{\citenamefont {Cui}\ \emph {et~al.}(2023)\citenamefont {Cui},
  \citenamefont {Peng}, \citenamefont {Zhang}, \citenamefont {Wei},
  \citenamefont {Yan},\ and\ \citenamefont {Chen}}]{PhysRevB.107.165414}%
  \BibitemOpen
  \bibfield  {author} {\bibinfo {author} {\bibfnamefont {Z.}~\bibnamefont
  {Cui}}, \bibinfo {author} {\bibfnamefont {M.}~\bibnamefont {Peng}}, \bibinfo
  {author} {\bibfnamefont {X.}~\bibnamefont {Zhang}}, \bibinfo {author}
  {\bibfnamefont {Q.}~\bibnamefont {Wei}}, \bibinfo {author} {\bibfnamefont
  {M.}~\bibnamefont {Yan}},\ and\ \bibinfo {author} {\bibfnamefont
  {G.}~\bibnamefont {Chen}},\ }\bibfield  {title} {\bibinfo {title}
  {Realization of multiple topological boundary states in phononic crystals},\
  }\href {https://doi.org/10.1103/PhysRevB.107.165414} {\bibfield  {journal}
  {\bibinfo  {journal} {Phys. Rev. B}\ }\textbf {\bibinfo {volume} {107}},\
  \bibinfo {pages} {165414} (\bibinfo {year} {2023})}\BibitemShut {NoStop}%
\bibitem [{\citenamefont {Hatsugai}(1993)}]{PhysRevLett.71.3697}%
  \BibitemOpen
  \bibfield  {author} {\bibinfo {author} {\bibfnamefont {Y.}~\bibnamefont
  {Hatsugai}},\ }\bibfield  {title} {\bibinfo {title} {Chern number and edge
  states in the integer quantum hall effect},\ }\href
  {https://doi.org/10.1103/PhysRevLett.71.3697} {\bibfield  {journal} {\bibinfo
   {journal} {Phys. Rev. Lett.}\ }\textbf {\bibinfo {volume} {71}},\ \bibinfo
  {pages} {3697} (\bibinfo {year} {1993})}\BibitemShut {NoStop}%
\bibitem [{\citenamefont {Koshino}\ and\ \citenamefont
  {McCann}(2009)}]{Koshino2009}%
  \BibitemOpen
  \bibfield  {author} {\bibinfo {author} {\bibfnamefont {M.}~\bibnamefont
  {Koshino}}\ and\ \bibinfo {author} {\bibfnamefont {E.}~\bibnamefont
  {McCann}},\ }\bibfield  {title} {\bibinfo {title} {Trigonal warping and
  berry's phase {$N\pi$} in {ABC}-stacked multilayer graphene},\ }\href
  {https://doi.org/10.1103/PhysRevB.80.165409} {\bibfield  {journal} {\bibinfo
  {journal} {Phys. Rev. B}\ }\textbf {\bibinfo {volume} {80}},\ \bibinfo
  {pages} {165409} (\bibinfo {year} {2009})}\BibitemShut {NoStop}%
\bibitem [{\citenamefont {Zak}(1989)}]{PhysRevLett.62.2747}%
  \BibitemOpen
  \bibfield  {author} {\bibinfo {author} {\bibfnamefont {J.}~\bibnamefont
  {Zak}},\ }\bibfield  {title} {\bibinfo {title} {Berry's phase for energy
  bands in solids},\ }\href {https://doi.org/10.1103/PhysRevLett.62.2747}
  {\bibfield  {journal} {\bibinfo  {journal} {Phys. Rev. Lett.}\ }\textbf
  {\bibinfo {volume} {62}},\ \bibinfo {pages} {2747} (\bibinfo {year}
  {1989})}\BibitemShut {NoStop}%
\end{thebibliography}%

\end{document}